# A goodness-of-fit test based on the empirical characteristic function and a comparison of tests for normality

J. Martin van Zyl

Department of Mathematical Statistics and Actuarial Science, University of the Free State, South Africa

**Abstract**  The normal distribution has the unique property that the cumulant generating function has only two terms, namely those involving the mean and the variance. Various tests based on the empirical characteristic function were proposed. In this work a simple normality test based on the studentized observations and the modulus of the empirical characteristic function is shown to outperform six of the most recognized test for normality in large samples.

## 1.  Introduction

A goodness-of-fit test statistic which is a function of the empirical characteristic function (ecf) and with an asymptotic standard normal distribution was derived by Murota and Takeuchi (1981). They derived a location and scale invariant test using studentized observations and showed that the use of a single value when calculating the ecf is sufficient to get good results with respect to power when testing hypotheses. Murota and Takeuchi (1981) compared their test against a test based on the sample kurtosis, and conducted a small simulation study, but did not compare the performance of their test against other tests for normality. In this work a test is proposed and the asymptotic distributional results derived from the results by Murota and Takeuchi (1981). A simulation study is conducted to compare the power of this test and six of the most recognized goodness-of-fit tests for normality. The test performs reasonably in small sample, but excellent in large samples with respect to power, and the test statistic is a simple normal test which will perform better as the sample increases.

Various overview simulation studies were conducted to investigate the performance of tests for normality. One of the most cited papers is the work by Yap and Sim (2011), but they did not include a goodness-of-fit test based on the empirical characteristic function. Their work is used as a guideline to decide which tests to include when looking at the performance of the test proposed in this work. The tests included are the Jarque–Bera, Shapiro-Wilk, Lilliefors, Anderson-Darling and D'Agostino and Pearson tests. A paper which included a very large selection of tests for normality is the work by Romao et al. (2013), but the test of Murota and Takeuchi (1981) was not included in this study.

Murota and Takeuchi (1981) proved that the square of the modulus of the empirical characteristic function converges weakly to a complex Gaussian process where the observations are standardized using affine invariant estimators of location and scale and they derived an expression for the asymptotic variance. Let $x_1,...,x_n$ be an i.i.d. sample of size $n$, from a distribution $F$. The characteristic function is $E(e^{itX}) = \phi(t)$ and it is estimated by the ecf

$$\hat{\phi}_F(t) = \frac{1}{n}\sum_{j=1}^{n} e^{itx_j},  \qquad (1)$$

The studentized sample is $z_1,...,z_n$, where $z_j = (x_j - \hat{\mu}_n)/\hat{\sigma}_n$, $j = 1,...,n$, with $\hat{\mu}_n = \overline{x}$ and $\hat{\sigma}_n^2 = S^2$. Denote the ecf, based on the standardized data, by $\hat{\phi}_S(t) = (1/n)\sum_{j=1}^{n} e^{itz_j}$.

The statistic proposed to test normality is

$$v_n(1) = \log(|\hat{\phi}_S(1)/\exp(-1/2)|), \qquad (2)$$

where $\sqrt{n}\,(v_n(1)) \sim N(0, 0.0431)$ asymptotically. In the simulation study it was found that the asymptotic normality approximation yields good results with respect to power for samples larger than $n=50$. A motivation for the ratio form of the statistic can be given in terms of cumulants. Thus reject normality if

$$|v_n(1)/(\sqrt{0.0431/n}| = |4.8168\sqrt{n}v_n(1)| > z_{1-\alpha/2}. \qquad (3)$$

The test can also be written as, reject normality if

$$|\sqrt{n}(\log(|\hat{\phi}_S(1)|) - (-1/2))/0.2076| = |\sqrt{n}(\log(|\hat{\phi}_S(1)|) + 1/2)/0.2076|$$

$$> z_{1-\alpha/2}.$$

A discussion of some of the tests where the characteristic is used can be found in the book by Ushakov (1999). A test based on the ecf which attracted attention and yielded good results is the test derived by Epps and Pulley (1983). This test is based on the expected value of the squared modulus of the difference between the ecf and the theoretical characteristic function under normality, with respect to a weight function which has a similar form as a normal density. Henze (1990) derived a large sample approximation for this test, but even the approximation is much more complicated than the test of Murota and Takeuchi (1981).

## 2  Methodology

### 2.1   Motivation for the proposed test statistic

A motivation will be given in terms of the cumulant generating function. The normal distribution has the unique property that the cumulant generating function cannot be a finite-order polynomial of degree larger than two, and the normal distribution is the only distribution for which all cumulants of order larger than 3 are zero (Cramér 1946, Lukacs 1972).

The motivation for the test will be shown by using the moment generating function, but experimentation showed that the use of the characteristic function rather than the moment generating function gives much better results when used to test for normality.
Consider a random variable $X$ with distribution $F$, mean $\mu$ and variance $\sigma^2$. The cumulant generating function $K_F(t)$ of $F$ can be written as $K_F(t) = \sum_{r=1}^{\infty} \kappa_r t^r / r!$, where $\kappa_r$ is the r-th cumulant. The first two cumulants are $\kappa_1 = E(X) = \mu$, $\kappa_2 = Var(X) = \sigma^2$. Since $K_F(t)$ is the logarithm of the moment generating function, the moment generating function can be written as $M_F(t) = E(e^{tX}) = e^{K_F(t)}$. It follows that

$$K_F(t) = \log(E(e^{tX}))$$

$$= \sum_{r=1}^{\infty} \kappa_r t^r / r!$$

$$= \left(\kappa_1 t + \kappa_2 t^2 / 2\right) + \left(\sum_{r=3}^{\infty} \kappa_r t^r / r!\right)$$

$$= \left(\mu t + \sigma^2 t^2 / 2\right) + \left(\sum_{r=3}^{\infty} \kappa_r t^r / r!\right).$$

Let $F_N$ denote a normal distribution with a mean $\mu$ and variance $\sigma^2$. $M_N(t)$ denotes the moment generating function of the normal distribution with cumulant generating function $K_N(t) = \mu t + \frac{1}{2}\sigma^2 t^2$. The logarithm of the ratio of the moment generating functions of $F$ and $F_N$ is given by

$$\log(M_F(t)/M_N(t)) = \log(\exp(K_F(t) - K_N(t)))$$

$$= [(\mu t + \tfrac{1}{2}\sigma^2 t^2) + \sum_{r=3}^{\infty} \kappa_r t^r / r!] - (\mu t + \tfrac{1}{2}\sigma^2 t^2)$$

$$= K_N(t) + \sum_{r=3}^{\infty} \kappa_r t^r / r! - K_N(t))$$

$$= \sum_{r=3}^{\infty} \kappa_r t^r / r!. \qquad (4)$$

If $F$ is a normal distribution, the sum given in (1) is zero. Replacing $K_F(t) - K_N(t)$ by $K_F(it) - K_N(it)$ it follows that

$$\log(|\log(\phi_F(t)) - \log(\phi_N(t))|) = \log(|K_F(it) - K_N(it)|)$$

$$= \sum_{r=0}^{\infty} \kappa_{4+2r} t^{4+2r} / (4+2r)!,$$

which would be equal to zero when the distribution $F$ is a normal distribution and this expression can be used to test for normality. Murota and Takeuchi (1981) used the fact that the square root of the log of the modulus of the characteristic function of a normal distribution is linear in terms of $t$, in other words $(-\log(|\phi_N(t)|^2))^{1/2}$ is a linear function of $t$.

## 2.2 Asymptotic variance

Let $\hat{\phi}_{NS}(t)$ denote the ecf calculated in the point $t$ using studentized normally distributed observations. An asymptotic variance of $v_n(t) = \log(|\hat{\phi}_{NS}(t)/\exp(-t^2/2)|)$ can be found by using the delta method and the results of Murota and Takeuchi (1981). They showed that the process defined by

$$\tilde{Z}(t) = \sqrt{n}(|\hat{\phi}_{NS}(t)|^2 - \exp(-t^2)), \qquad (5)$$

converges weakly to a zero mean Gaussian process and variance

$$E(\tilde{Z}^2(t)) = 4\exp(-2t^2)(\cosh(t^2) - 1 - t^4/2). \qquad (6)$$

Note that $\hat{\phi}_{NS}(t) = e^{-t^2/2}$ and by applying the delta method it follows that

$$Var(|\hat{\phi}_{NS}(t)|^2) \approx Var(|\hat{\phi}_{NS}(t)|)(2|\hat{\phi}_{NS}(t)|),^2$$

thus $Var(|\hat{\phi}_{NS}(t)|) \approx Var(|\hat{\phi}_{NS}(t)|^2)/(2|\hat{\phi}_{NS}(t)|)^2$.

By applying the delta method again it follows that

$$\begin{aligned}
Var(v(t)) &= Var(\log(|\hat{\phi}_{NS}(t)/e^{-1/2}|)) \\
&\approx (1/|\hat{\phi}_{NS}(t)|^2)Var(|\hat{\phi}_{NS}(t)|) \\
&= Var(|\hat{\phi}_{NS}(t)|^2)/4(|\hat{\phi}_{NS}(t)|)^4 \\
&= (\cosh(t^2) - 1 - t^4/2)/n. \qquad (7)
\end{aligned}$$

The statistic $v_n(t)$ converges weakly to a Gaussian distribution with mean zero and variance $Var(v_n(t)) = (\cosh(t^2) - 1 - t^4/2)/n$, and

$$Var(v_n(1)) = 0.0431/n, \ t = 1. \qquad (8)$$

The test used is: $|v_n(1)/(\sqrt{0.0431/n}| > z_{1-\alpha/2}$. In the following figure the average of the log the modulus calculated in the point, $t = 1$, using standard normally distributed samples, for various samples sizes is shown. The solid line is where studentized observations were used and the dashed line where the ecf is calculated using the original sample. It can be seen that there is a large bias in small samples and the studentized ecf has less variation.

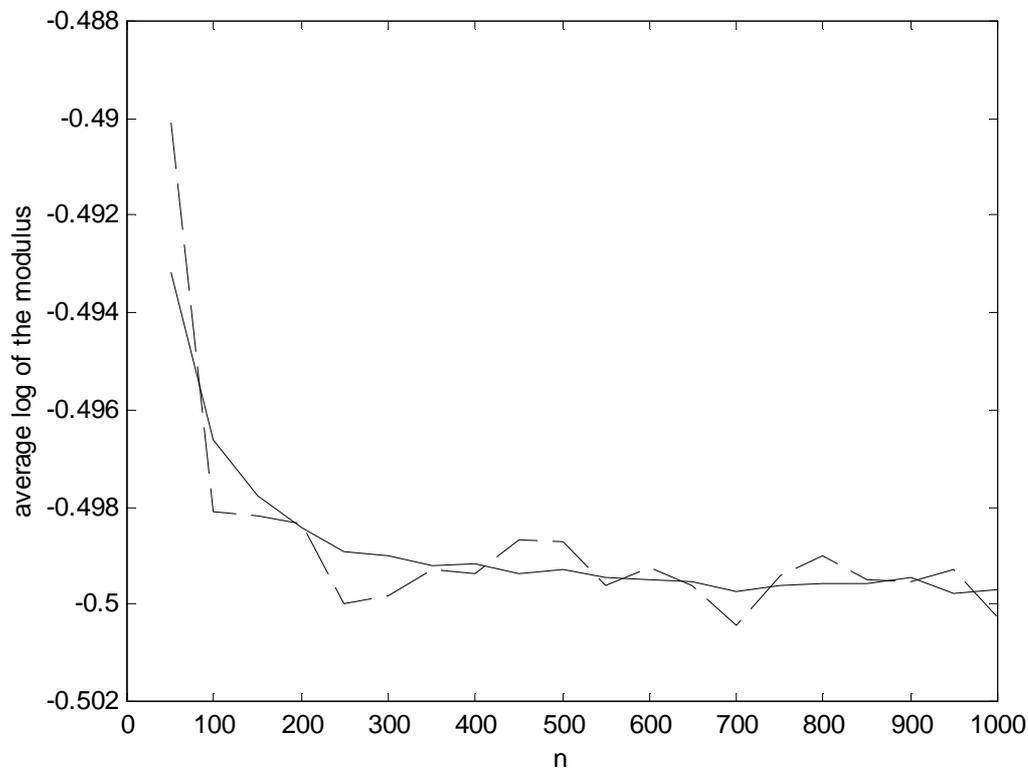

**Fig. 1** Plot of the average log of the modulus of the ecf for various sample sizes calculated using $m = 5000$ calculated using samples form a standard normal distribution. The solid line is where studentized observations are used and the dashed line using the original sample. Calculated in the point $t=1$, and the expected value is -0.5.

In the following histogram 5000 simulated values of

$v_n(1)/(\sqrt{0.0431/n}\,| = 4.8168\sqrt{n}\,v_n(1)$ are shown, where the $v_n(1)$'s are calculated using simulated samples of size $n=1000$ from a standard normal distribution.

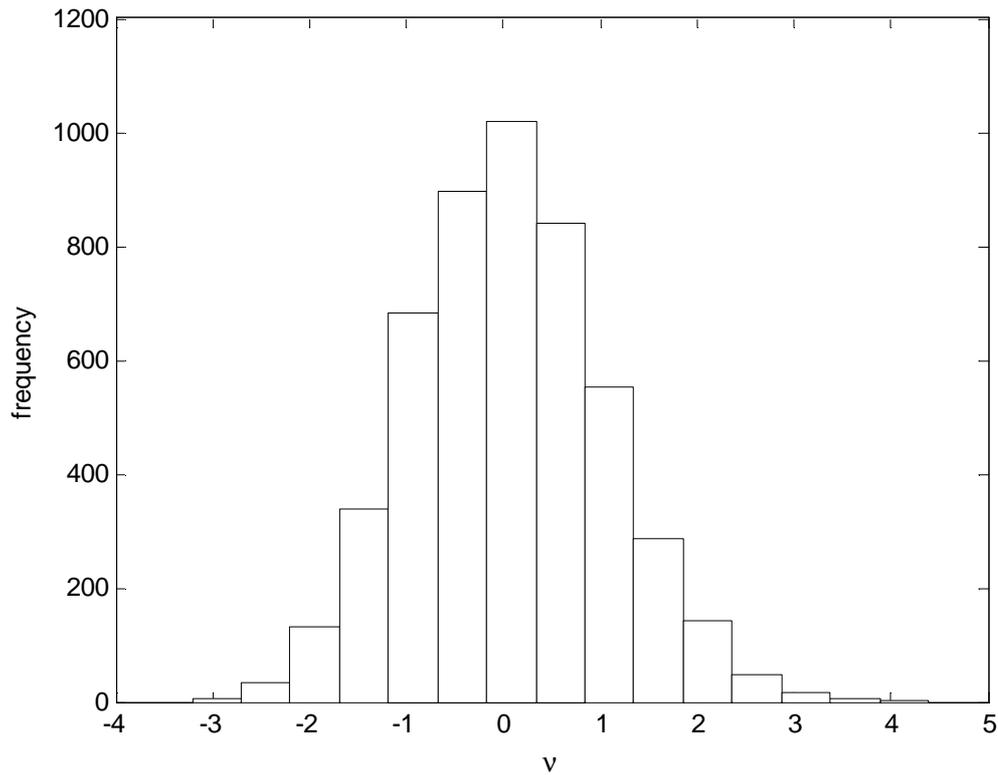

**Fig. 2** Histogram of $m=5000$ simulated values of $v_n(1)/(\sqrt{0.0431/n})$, with $n=1000$. Calculated using normally distributed samples, data standardized using estimated parameters.

In Figure 3 the variance of $v(1)$ is estimated for various sample sizes, based on 1000 estimated values of $v(1)$ for each sample size considered. The estimated variances is plotted against the asymptotic variance $Var(v_n) = 0.0431/n$.

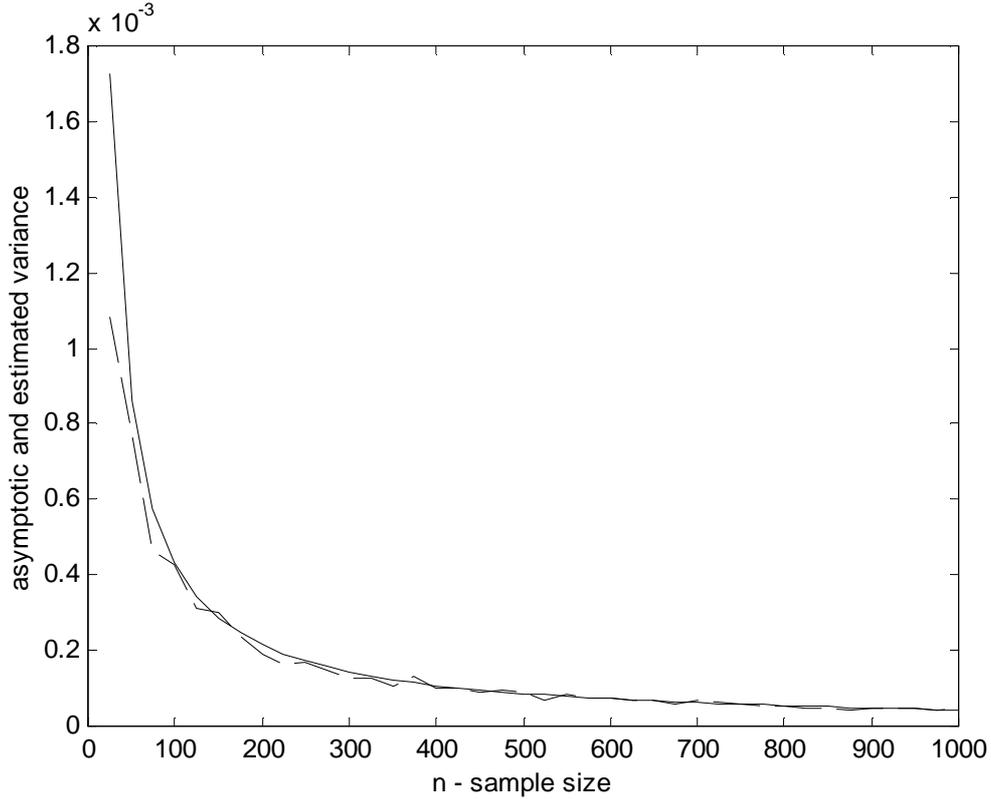

**Fig. 3** Estimated variance of $v_n$ and asymptotic variance for various values of *n*. Dashed line the estimated variance. Estimated variance calculated using 1000 normally distributed samples, data standardized using estimated parameters.

## 3. Simulation study

The paper of Yap and Sim (2011) is used as a guideline to compare the proposed test against other tests for normality. The proposed test will be denoted by ECFT in the tables. The power of the test will be compared against several tests for normality:

- The Lilliefors test (LL), Lilliefors (1967) which is a slight modification of the Kolmogorov-Smirnov test for where parameters are estimated.

- The Jarque-Bera test (JB), Jarque and Bera (1987), where the skewness and kurtosis is combined to form a test statistics.

- The Shapiro-Wilks test (SW), Shapiro and Wilk (1965). This test makes use of properties of order statistics and were later developed to be used for large samples too by Royston (1992).
- The Anderson-Darling test (AD), Anderson and Darling .
- The D'Agostino and Pearson test (DP), D'Agostino and Pearson (1973). This statistic combines the skewness and kurtosis to check for deviations from normality.

Samples are generated from a few symmetric unimodal distributions with sizes $n = 30, 50, 100, 250, 500, 750, 1000$. The proportion rejections are reported based on $m$=5000 repetitions. The test are conducted at the 5% level and for the ecf, the normal approximation is used.

The following symmetric distributions are considered, uniform on the interval [0,1], the logistic distribution with mean zero the standard t-distribution and the Laplace distribution with mean zero and scale parameter one. The standard $t$-distribution with 4, 10 and 15 degrees of freedom. Skewed distributions and multimodal distributions would not be investigated, since in large samples such samples can be already excluded with certainty as being not from a normal distribution by looking at the histograms.

All the tests performed for the ecf test were conducted using the normal approximation, but percentiles can also easily be simulated. Simulated estimates of the Type I error for $n = 30, 50, 100, 250, 500, 750, 1000$, are given in Table 1 based on $m = 5000$ simulated samples. The simulated samples are standard normally distributed and studentized to calculate the Type I error.

| n | Type I error Normality assumption | LL | JB | SW | AD | DP |
|---|---|---|---|---|---|---|
| 30 | 0.0438 | 0.0534 | 0.0512 | 0.0580 | 0.0506 | 0.0552 |
| 50 | 0.0466 | 0.0480 | 0.0514 | 0.0532 | 0.0494 | 0.0570 |
| 100 | 0.0442 | 0.0504 | 0.0460 | 0.0444 | 0.0468 | 0.0500 |
| 250 | 0.0442 | 0.0524 | 0.0506 | 0.0452 | 0.0516 | 0.0552 |
| 500 | 0.0488 | 0.0498 | 0.0490 | 0.0430 | 0.0480 | 0.0540 |
| 750 | 0.0498 | 0.0460 | 0.0480 | 0.0414 | 0.0488 | 0.0526 |
| 1000 | 0.0496 | 0.0500 | 0.0510 | 0.0404 | 0.0530 | 0.0544 |

**Table 1** Simulated percentiles to test for normality at the 5% level. Calculated from $m=5000$ simulated samples of size $n$ each in the point $t=1$.

In Table 2 the rejection rates, when testing at the 5% level and symmetric distributions, are shown for various sample sizes based on 10000 samples each time.

In table 1 the t-distribution where not all moments exist is considered. The JB, DP and ECFT tests performs best, and in large samples the ECFT test performs the best.

|       | n    | ECFT   | LL     | JB     | SW     | AD     | DP     |
|-------|------|--------|--------|--------|--------|--------|--------|
| t(4)  | 50   | 0.5798 | 0.3066 | 0.5510 | 0.4778 | 0.4302 | 0.5214 |
|       | 100  | 0.8144 | 0.4896 | 0.7810 | 0.7152 | 0.6514 | 0.7340 |
|       | 250  | 0.9836 | 0.8354 | 0.9754 | 0.9606 | 0.9406 | 0.9648 |
|       | 500  | 0.9998 | 0.9866 | 0.9998 | 0.9992 | 0.9974 | 0.9990 |
|       | 750  | 1.0000 | 0.9990 | 1.0000 | 1.0000 | 1.0000 | 1.0000 |
|       | 1000 | 1.0000 | 0.9998 | 1.0000 | 1.0000 | 1.0000 | 1.0000 |
| t(10) | 50   | 0.1896 | 0.0886 | 0.1900 | 0.1446 | 0.1184 | 0.1768 |
|       | 100  | 0.3194 | 0.1100 | 0.3098 | 0.2252 | 0.1614 | 0.2722 |
|       | 250  | 0.5604 | 0.1766 | 0.5240 | 0.4052 | 0.2802 | 0.4644 |
|       | 500  | 0.8078 | 0.2852 | 0.7572 | 0.6376 | 0.4838 | 0.7032 |
|       | 750  | 0.9110 | 0.3928 | 0.8692 | 0.7866 | 0.6396 | 0.8336 |
|       | 1000 | 0.9632 | 0.5136 | 0.9400 | 0.8824 | 0.7688 | 0.9170 |
| t(15) | 50   | 0.1376 | 0.0612 | 0.1418 | 0.1050 | 0.0762 | 0.1358 |
|       | 100  | 0.2114 | 0.0794 | 0.2028 | 0.1408 | 0.1078 | 0.1778 |
|       | 250  | 0.3344 | 0.0934 | 0.3110 | 0.2160 | 0.1384 | 0.2694 |
|       | 500  | 0.5182 | 0.1402 | 0.4834 | 0.3622 | 0.2318 | 0.4212 |
|       | 750  | 0.6558 | 0.1694 | 0.6038 | 0.4704 | 0.3122 | 0.5460 |
|       | 1000 | 0.7712 | 0.2180 | 0.7184 | 0.5868 | 0.4004 | 0.6672 |

**Table 2** Simulated power of normality tests. Rejection proportions when testing for normality at the 5% level.

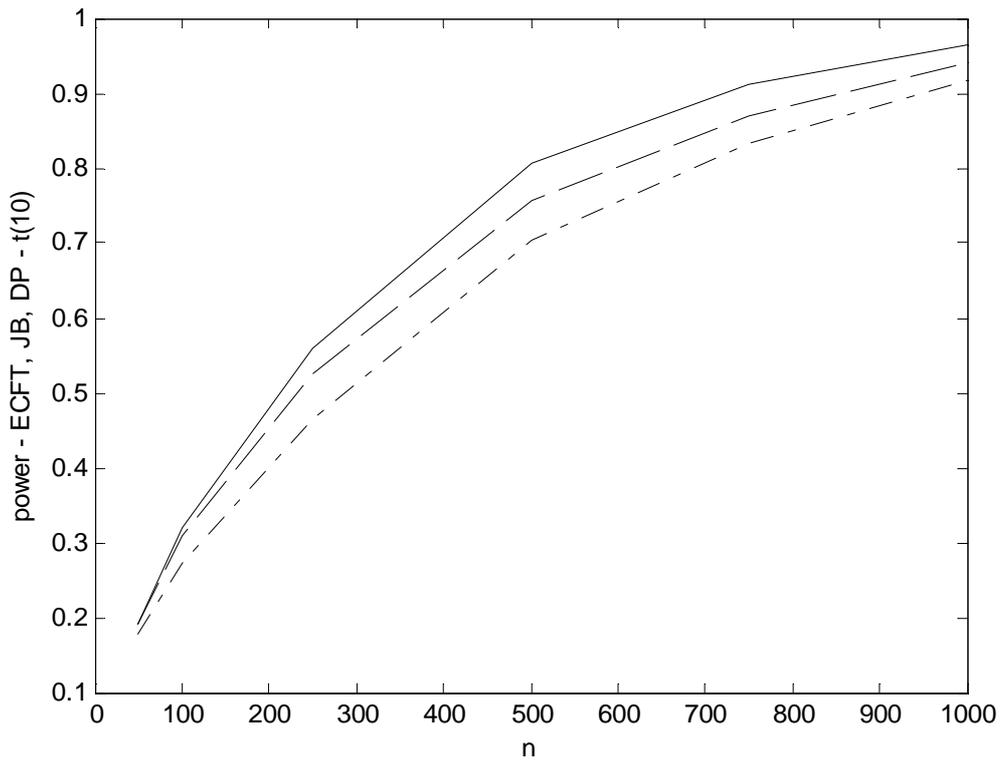

**Fig. 4** Plot of the three best performing tests with respect to power, testing for normality, data t-distributed with 10 degrees of freedom. Solid line, ecf test, dashed line JB test and dash-dot the DP test.

|  | n | ECFT | LL | JB | SW | AD | DP |
|---|---|---|---|---|---|---|---|
| U(0,1) | 50 | 0.1360 | 0.2626 | 0.0096 | 0.5772 | 0.5778 | 0.7952 |
|  | 100 | 0.9510 | 0.5910 | 0.7410 | 0.9870 | 0.9522 | 0.9976 |
|  | 250 | 1.0000 | 0.9866 | 1.0000 | 1.0000 | 1.0000 | 1.0000 |
|  | 500 | 1.0000 | 1.0000 | 1.0000 | 1.0000 | 1.0000 | 1.0000 |
| Laplace | 50 | 0.6196 | 0.4398 | 0.5660 | 0.5164 | 0.5482 | 0.5138 |
|  | 100 | 0.8686 | 0.7008 | 0.8026 | 0.7888 | 0.8238 | 0.7448 |
|  | 250 | 0.9958 | 0.9762 | 0.9848 | 0.9884 | 0.9942 | 0.9728 |
|  | 500 | 1.0000 | 1.0000 | 1.0000 | 1.0000 | 1.0000 | 0.9998 |
| Logistic | 50 | 0.2758 | 0.1192 | 0.2656 | 0.1976 | 0.1716 | 0.2416 |
|  | 100 | 0.4240 | 0.1542 | 0.3932 | 0.2960 | 0.2416 | 0.3428 |
|  | 250 | 0.7150 | 0.2748 | 0.6432 | 0.5416 | 0.4436 | 0.5770 |
|  | 500 | 0.9338 | 0.5068 | 0.8930 | 0.8220 | 0.7480 | 0.8506 |
|  | 750 | 0.9842 | 0.6958 | 0.9670 | 0.9370 | 0.9028 | 0.9504 |
|  | 1000 | 0.9978 | 0.8280 | 0.9904 | 0.9820 | 0.9626 | 0.9872 |

**Table 3** Simulated power of normality tests. Rejection proportions when testing for normality at the 5% level.

It can be seen the ECFT outperforms the other tests with respect to power in large samples, especially when testing data from a logistic distribution.

Samples were simulated from a mixture of two normal distributions, with a proportion $\alpha$ from a standard normal and a proportion $1-\alpha$ from a normal with variance $\sigma^2$. This can also be considered as a contaminated distribution. The results are shown in Table 4. The proposed test yielded good results.

| $(\sigma,\alpha)$ | n | ECFT | LL | JB | SW | AD | DP |
|---|---|---|---|---|---|---|---|
| (2,0.2) | 30 | 0.2518 | 0.1056 | 0.2580 | 0.1952 | 0.1596 | 0.2478 |
| | 50 | 0.3832 | 0.1378 | 0.3676 | 0.2792 | 0.2134 | 0.3398 |
| | 100 | 0.5986 | 0.2016 | 0.5640 | 0.4484 | 0.3344 | 0.5016 |
| | 250 | 0.8834 | 0.3630 | 0.8506 | 0.7646 | 0.6200 | 0.8016 |
| | 500 | 0.9872 | 0.6378 | 0.9784 | 0.9546 | 0.8790 | 0.9684 |
| | 750 | 0.9980 | 0.8238 | 0.9978 | 0.9934 | 0.9746 | 0.9958 |
| | 1000 | 1.0000 | 0.9242 | 1.0000 | 0.9998 | 0.9948 | 1.0000 |
| (0.5,0.2) | 30 | 0.0884 | 0.0676 | 0.0940 | 0.0816 | 0.0752 | 0.0918 |
| | 50 | 0.1120 | 0.0710 | 0.1074 | 0.0868 | 0.0814 | 0.0984 |
| | 100 | 0.1634 | 0.0894 | 0.1406 | 0.0994 | 0.1100 | 0.1166 |
| | 250 | 0.2784 | 0.1446 | 0.2084 | 0.1492 | 0.2050 | 0.1652 |
| | 500 | 0.4406 | 0.2448 | 0.3220 | 0.2380 | 0.3532 | 0.2552 |
| | 750 | 0.5766 | 0.3600 | 0.4282 | 0.3622 | 0.5004 | 0.3566 |
| | 1000 | 0.7040 | 0.4646 | 0.5398 | 0.4678 | 0.6346 | 0.4602 |
| (2.0,0.5) | 30 | 0.2034 | 0.1066 | 0.1972 | 0.1562 | 0.1428 | 0.1896 |
| | 50 | 0.2928 | 0.1326 | 0.2634 | 0.1970 | 0.1946 | 0.2358 |
| | 100 | 0.4954 | 0.2136 | 0.4178 | 0.3322 | 0.3292 | 0.3538 |
| | 250 | 0.8244 | 0.4572 | 0.7160 | 0.6416 | 0.6696 | 0.6200 |
| | 500 | 0.9788 | 0.7760 | 0.9420 | 0.9226 | 0.9372 | 0.9062 |
| | 750 | 0.9988 | 0.9230 | 0.9888 | 0.9846 | 0.9876 | 0.9796 |
| | 1000 | 1.0000 | 0.9838 | 0.9992 | 0.9986 | 0.9990 | 0.9972 |

**Table 4** Simulated power of normality tests. Rejection proportions when testing for normality at the 5% level. Mixture of two normal distributions (contaminated data)

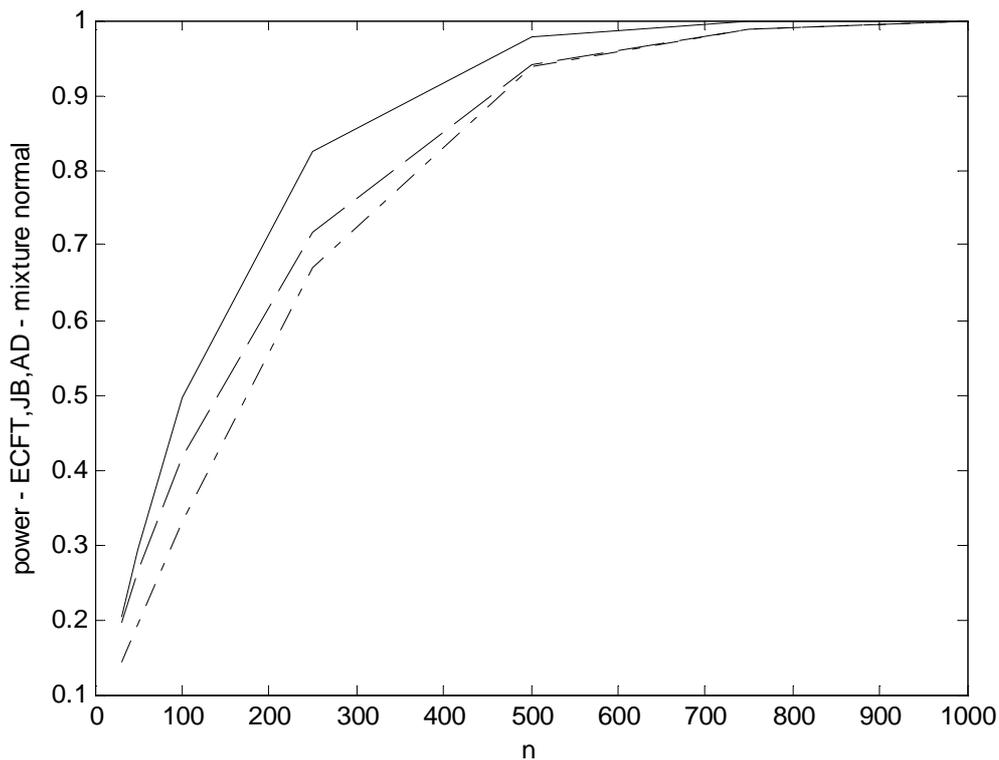

**Fig 5** Plot of the three best performing tests with respect to power, testing for normality, data mixture of normal distributions with two components. Mixture .5N(0,1)+0.5N(0,0,16). Solid line, ecf test, dashed line JB test and dash-dot the AD test.

## 4 Conclusions

The proposed test performs better with respect to power in large samples than the other tests for normality for the distributions considered in the simulation study . In small samples of say less than $n = 50$, it was found that the test of D'Agostino and Pearson (1973) was often either the best performing or close to the best performing test.

In practice one would not test data from a skewed distribution for normality in large samples. The simple normal test approximation will perform better, the larger the sample is. The asymptotic normality and variance properties, which is of a very simple form, can be used in large samples. This test can be recommended as probably the test of choice in terms of power and easy of application in large samples.